# High-fidelity readout and control of a nuclear spin qubit in silicon

Jarryd J. Pla[1], Kuan Y. Tan[1†], Juan P. Dehollain[1], Wee H. Lim[1†], John J. L. Morton[2], Floris A. Zwanenburg[1†], David N. Jamieson[3], Andrew S. Dzurak[1], Andrea Morello[1]

[1]Centre for Quantum Computation and Communication Technology, School of Electrical Engineering & Telecommunications, University of New South Wales, Sydney, New South Wales 2052, Australia.

[2]London Centre for Nanotechnology, University College London, London WC1H 0AH, UK.

[3]Centre for Quantum Computation and Communication Technology, School of Physics, University of Melbourne, Melbourne, Victoria 3010, Australia.

[†]Present addresses: Department of Applied Physics/COMP, Aalto University, P.O. Box 13500, FI-00076 AALTO, Finland (K.Y.T.); Asia Pacific University of Technology and Innovation, Technology Park Malaysia, Bukit Jalil, 57000 Kuala Lumpur, Malaysia (W.H.L.); NanoElectronics Group, MESA+ Institute for Nanotechnology, University of Twente, Enschede, The Netherlands (F.A.Z.).

*A single nuclear spin holds the promise of being a long-lived quantum bit or quantum memory, with the high fidelities required for fault-tolerant quantum computing. We show here that such promise could be fulfilled by a single phosphorus ($^{31}$P) nuclear spin in a silicon nanostructure. By integrating single-shot readout of the electron spin with on-chip electron spin resonance, we demonstrate the quantum non-demolition, electrical single-shot readout of the nuclear spin, with readout fidelity better than 99.8% - the highest for any solid-state qubit. The single nuclear spin is then operated as a qubit by applying coherent radiofrequency (RF) pulses. For an ionized $^{31}$P donor we find a nuclear spin coherence time of 60 ms and a 1-qubit gate control fidelity exceeding 98%. These results demonstrate that the dominant technology of modern electronics can be adapted to host a complete electrical measurement and control platform for nuclear spin-based quantum information processing.*

Quantum computers have the potential to revolutionize aspects of modern society from fundamental science and medical research[1-3] to data analysis[4,5]. The successful demonstration of such a machine depends on the ability to perform high-fidelity control and measurement of individual qubits[6] – the building blocks of a quantum computer. Errors introduced by quantum operations and measurements can be mitigated by employing quantum error correction protocols[7], provided that the probabilities of the errors occurring are below certain stringent thresholds[8-10]. To date, the state-of-the-art in high-fidelity qubit control and readout has been defined by laser cooled atoms in electromagnetic traps[11,12] - a result made possible because of their extreme isolation in a near perfect vacuum.

Qubits based on physical systems in the solid-state[13-19] are attractive because of their potential for scalability using modern integrated circuit fabrication technologies[20]. However, they tend to exhibit much lower system fidelities owing to interactions with their host environment[21]. An ability to combine the control and measurement fidelities of trapped atoms with the scalability benefits inherent to solid-state implementations is therefore highly desirable. The nuclear spin of a single atom is a promising candidate in this regard; it represents a simple, well-isolated quantum system. It can be oriented and caused to precess using combinations of static and oscillating magnetic fields[22]. The spin orientation persists for a very long time, even when the nucleus is hosted by a crystal[23] or a molecule[24]. This property has been exploited for a variety of applications, ranging from magnetic resonance imaging (MRI)[25] to the execution of quantum algorithms[26].

One of the earliest proposals for quantum computing in the solid-state advocated the use of the nuclear spin of individual $^{31}$P dopant atoms in silicon to encode and process quantum information[27]. Silicon is an ideal platform for spin-based quantum information processing because it can be enriched in the nuclear spin-zero $^{28}$Si isotope[28], providing an effective "semiconductor vacuum" and very long spin coherence times[29,30]. Experiments in bulk phosphorus-doped isotropically-enriched silicon ($^{28}$Si:P) have already highlighted the potential of this system, where the $^{31}$P nuclear spin has been implemented as a quantum memory[31] and as a qubit with extraordinary coherence lifetimes > 180 s (ref. 32). However, due to detection limitations, experiments have so far only been carried out on large ensembles of $^{31}$P nuclei, typically several billion in number[32,33]. To realize nuclear spin-based solid-state quantum computing, one must first isolate, measure and control individual nuclear spins.

Here we demonstrate the readout and coherent manipulation of a single $^{31}$P nuclear spin qubit in a silicon chip. As with modern microelectronic circuits, our qubit operations are performed electrically through the application of on-chip voltage and current signals. We show that this solid-state system is capable of realizing fidelities approaching those in vacuum-based ion-trap qubits, raising the prospects for scalable and fault-tolerant quantum computation in silicon.

**Electrical single-shot nuclear spin readout**

Measuring the state of a single nuclear spin is challenging due to its weak magnetic moment. In the solid-state, this has only been achieved in the nitrogen-vacancy centre in diamond[34], with optical detection, and on a rare earth terbium ion by performing electrical transport measurements through a single molecule[24]. In both of these cases, a coupled electron spin was used to read out the nuclear spin, and thus the ability to measure a single electron spin was prerequisite.

One of the most versatile and successful methods of electrically measuring single electron spins in the solid-state relies on a process known as spin-to-charge conversion[35]. The electron is displaced to a different location depending on its spin state, and the resulting change in the local potential can be detected using a nano-scale electrometer. This readout method has been applied to the electron spin of a $^{31}$P donor in silicon[19,36], and forms the basis of our nuclear spin readout.

We employ an on-chip all-electrical detection method, based on nanostructures that are compatible with silicon metal-oxide-semiconductor (Si MOS) fabrication standards. Both the electron and the nuclear spins are read out electrically using a compact nano-scale device[37] (Fig. 1a) consisting of ion-implanted phosphorus donors[38], tunnel-coupled to a silicon MOS single-electron transistor (SET)[39]. To achieve the single-shot readout of a $^{31}$P nuclear spin, we implement a three-stage process (Fig. 1d) which amplifies the nuclear spin state energy splitting (~ 100 neV) by a factor of order $10^9$. In the first stage, we exploit electron spin resonance (ESR)[19] to map the nuclear spin state onto the electron spin. The second stage involves performing spin-to-charge conversion[36] which projects the electron spin state onto a charge state of the $^{31}$P donor. Finally, we utilize the SET to provide a current, dependent upon the donor charge state, with a total signal energy of ~ 100 eV under typical experimental conditions.

The $^{31}$P donor in silicon can be thought of as the equivalent of a hydrogen atom, in a solid-state matrix. It possesses a nuclear spin $I = 1/2$, and the excess charge of the phosphorus nucleus, as compared to the surrounding silicon nuclei, creates a Coulomb potential that can bind an extra electron (with spin $S = 1/2$) in the neutral $D^0$ donor charge state. Therefore, a single $^{31}$P donor constitutes a 2-qubit system, where the two qubits interact with an external magnetic field $B_0$ in proportion to their gyromagnetic ratios: $\gamma_n = 17.23$ MHz/T for the nucleus[40]; and $\gamma_e = g\mu_B/h = 27.97$ GHz/T for the electron, where $\mu_B$ is the Bohr magneton and $g = 1.9985$ (ref. 41) is the Landé $g$-factor. In addition, they interact with each other through the hyperfine interaction $A = 117.53$ MHz (ref. 41), which arises from the overlap of the electron wavefunction with that of the $^{31}$P nucleus. If $\gamma_e B_0 \gg A > 2\gamma_n B_0$, the eigenstates of the two-spin system are approximately (in ascending order of energy) $|\downarrow\Uparrow\rangle$, $|\downarrow\Downarrow\rangle$, $|\uparrow\Downarrow\rangle$, $|\uparrow\Uparrow\rangle$, where the thin

(thick) arrow indicates the orientation of the electron (nuclear) spin (Fig. 1b). The system can be transformed to that of a single $^{31}$P nuclear spin (Fig. 1c) with eigenstates $|\Uparrow\rangle$ and $|\Downarrow\rangle$, by utilizing the nanostructure device (Fig. 1a) to ionize the donor to the D$^+$ charge state.

The $^{31}$P nuclear spin readout experiment begins by performing ESR on its bound donor electron[19], which is manipulated using a resonant microwave excitation and measured in a single shot using an adjacent SET[36]. The microwave pulses are delivered by a custom designed, on-chip broadband planar transmission line, terminated with a short-circuit ~100 nm away from the location of the donor[42] (see Supplementary Information for details). The hyperfine coupling $A$ produces an effective magnetic field on the electron, which can add to or subtract from the external field $B_0$ depending on the orientation of the nuclear spin. Therefore, the system exhibits two possible ESR frequencies: $\nu_{e1} \approx \gamma_e B_0 - A/2$ for nuclear spin $|\Downarrow\rangle$; and $\nu_{e2} \approx \gamma_e B_0 + A/2$ for nuclear spin $|\Uparrow\rangle$. These expressions become exact in the limit $\gamma_e B_0 \gg A$, with deviations at low magnetic fields[40]. In a single-atom experiment, if we assume the ESR measurement duration to be much shorter than the nuclear spin flip time, then we expect only one active ESR frequency at any instant. Detecting electron spin resonance at the frequency $\nu_{e1}$ therefore indicates that the nuclear spin is in state $|\Downarrow\rangle$, whereas detection at $\nu_{e2}$ implies the nuclear spin is $|\Uparrow\rangle$.

Having identified the two resonance frequencies through an ESR experiment (see Fig. 1e and also ref. 19), we performed repeated measurements of the nuclear spin state (Fig. 2a) by toggling the microwave frequency $\nu_{ESR}$ between $\nu_{e1}$ and $\nu_{e2}$, averaging 250 electron spin measurements at each point (acquisition time 260 ms) to obtain the electron spin-up fraction $f_\uparrow$. In order to maximize the probability of flipping the electron spin in each shot, we execute a fast adiabatic passage by applying a frequency chirp centered about the ESR transition[43]. If the quantity $\Delta f_\uparrow = f_\uparrow(\nu_{e2}) - f_\uparrow(\nu_{e1})$ is positive, we assign the nuclear state $|\Uparrow\rangle$, and vice versa. A histogram of $\Delta f_\uparrow$ (Fig. 2d) shows two well-separated Gaussian peaks, corresponding to the two possible nuclear orientations. The widths of the peaks result from a combination of effects including: thermal broadening (caused by microwave induced heating), charge fluctuations (which alter the device biasing) and an imperfect adiabatic passage. These effects act to reduce the signal-to-noise ratio (SNR) of the measurement. The nuclear spin readout error (Fig. 2e) is obtained by fitting the two peaks and integrating each Gaussian beyond a discrimination threshold $\Delta f_{th}$. At the optimal value of $\Delta f_{th} = -0.025$, the SNR-limited readout error is $2\times10^{-7}$. A further analysis of measurement errors is presented later in the section discussing qubit fidelities.

**Quantum jumps of the nuclear spin state**

We observe that the nuclear spin state remains unchanged for several minutes before exhibiting a "quantum jump" to the opposite state[34]. From Fig. 2b, it is evident that the nuclear spin is predominantly polarized in the $|\Uparrow\rangle$ state. We attribute this phenomenon to an electron-nuclear spin flip-flop process, in which the energy difference $E_{\uparrow\Downarrow} - E_{\downarrow\Uparrow}$ (i.e. between states $|\uparrow\Downarrow\rangle$ and $|\downarrow\Uparrow\rangle$) is released to the phonon bath. The spin-phonon coupling may arise from the modulation of the hyperfine coupling caused by lattice deformation[44]. The same mechanism was invoked to explain the ~100 s decay time of spin polarization stored in a $^{31}$P ensemble[33]. Since $E_{\uparrow\Downarrow} - E_{\downarrow\Uparrow} >> k_B T$ in our experiment, this process acts only in the direction $|\uparrow\Downarrow\rangle \rightarrow |\downarrow\Uparrow\rangle$ (i.e., only spontaneous emission of phonons occurs), and should not be responsible for the observed nuclear spin jumps from $|\Uparrow\rangle$ to $|\Downarrow\rangle$.

We have established the cause of the $|\Uparrow\rangle \rightarrow |\Downarrow\rangle$ transition by modifying the readout pulse protocol to include a resonant tunneling phase, during which random tunneling of $|\downarrow\rangle$ electrons back and forth between the $^{31}$P donor and the SET island can occur (see Supplementary Information). This process can be viewed as a modulation of the hyperfine interaction between the electron and nuclear spins with dynamics governed by the tunneling electron, and is observed to decrease the $|\Uparrow\rangle$ state lifetime (Fig. 2c).

In Fig. 2f we plot the lifetime of the nuclear $|\Downarrow\rangle$ and $|\Uparrow\rangle$ states as a function of the rate of donor ionization/neutralization $\Gamma_{\text{ion/neut}}$. We find that the lifetime of the nuclear $|\Downarrow\rangle$ is approximately independent of $\Gamma_{\text{ion/neut}}$, as expected if the process is dominated by electron-nuclear spin flip-flops with phonon emission. Conversely, the lifetime of the nuclear $|\Uparrow\rangle$ is longer and inversely proportional to $\Gamma_{\text{ion/neut}}$. The results in Fig. 2f are accurately reproduced by a simulation where the hyperfine coupling is modulated in a random process that replicates the electron tunneling times as extracted from the measurements (see Supplementary Information for details).

**Coherent control: Single-spin nuclear magnetic resonance and Rabi oscillations**

By exploiting the broadband nature of our on-chip microwave transmission line, we perform a nuclear magnetic resonance (NMR) experiment on the $^{31}$P nuclear spin (Fig. 3). We expect two NMR frequencies depending on the state of the electron: $\nu_{n1} = A/2 + \gamma_n B_0$ when the electron spin is $|\downarrow\rangle$; and $\nu_{n2} = A/2 - \gamma_n B_0$ when the electron spin is $|\uparrow\rangle$ (Fig. 1b). The nuclear resonance is detected by measuring the absolute difference in electron spin-up counts between the two ESR frequencies, $|\Delta f_\uparrow| = |f_\uparrow(\nu_{e2}) - f_\uparrow(\nu_{e1})|$, as a function of the NMR frequency $\nu_{\text{NMR}}$. Off-resonance, we find the normal value $|\Delta f_\uparrow| \approx$ 0.4, as observed in the nuclear spin readout experiments (Fig. 2b), because the nucleus retains its spin state for a very long time. Conversely, an 8 ms long resonant excitation quickly randomizes the nuclear spin state, causing $|\Delta f\uparrow|$ to drop towards zero. $\nu_{n1}$ is found by applying an NMR burst before the ESR excitation (Fig. 3a), whereas for $\nu_{n2}$ we swap the order of ESR and NMR, to achieve a higher probability of having the electron spin $|\uparrow\rangle$, as required to observe the $\nu_{n2}$ resonance (Fig. 3b).

Since we have full control over the charge state of the donor, we can also perform an NMR experiment while the donor is ionized (Fig. 3c), as recently demonstrated in a bulk Si:P sample[45]. In this case there is only one resonance frequency, $\nu_{n0} = \gamma_n B_0$. The electron is placed back onto the donor after the NMR burst, for the purpose of reading out the nuclear spin state. Fig. 3d shows the magnetic field dependence of the three NMR frequencies, which agree with the expected values assuming the bulk $^{31}$P gyromagnetic ratio $\gamma_n = 17.23$ MHz/T (ref. 40). This observation confirms that the system under study is indeed a single $^{31}$P phosphorus atom. Furthermore, from these measurements we find that $g = 1.9987(6)$ (see Supplementary Information), within ~ 0.01% of the bulk value for Si:P, whereas the hyperfine splitting $A = 114.30(1)$ MHz is close to, but not identical with, the bulk value of 117.52 MHz (ref. 41). We interpret this as evidence of a Stark shift of the hyperfine coupling[46], caused by a distortion of the electron wavefunction under the strong electric fields present in the gated nanostructure. This observation is important because the Stark shift of $A$ was proposed by Kane[27] as a mechanism to address individual $^{31}$P nuclear spin qubits while applying a global microwave field.

By applying the NMR excitation for a much shorter duration, we are able to produce coherent superpositions of the nuclear spin states. In the Bloch sphere representation, where the states $|\Uparrow\rangle$ and $|\Downarrow\rangle$ reside at the poles, the short RF excitation produces a controlled rotation about the X or Y axis. These rotations can be observed by varying the length of the RF burst, which produces the Rabi oscillations of Fig. 4.

For the neutral ($D^0$) donor, we first initialize the electron in the $|\downarrow\rangle$ state. A pulse of length $t_p$ and at the $\nu_{n1}$ resonance – as determined from the NMR spectroscopy above - is applied immediately after, followed by a single-shot readout of the nuclear spin state (see Fig. 4a). A total of 200 measurements are performed at each $t_p$, and the induced nuclear spin flip probability $P_n$ is found. The result is the coherent Rabi oscillations of Fig. 4b, whose frequency $f_{rabi}$ scales linearly with RF excitation amplitude $P_{NMR}^{1/2}$ (Fig. 4c). The visibility of the oscillations in Fig. 4b is ~ 60%. Deviations from ideality are most likely due to erroneous initialization in the $|\uparrow\rangle$ state[19] caused by the non-zero electron temperature. The maximum Rabi frequency attained in our experiments was ~ 20 kHz. Stray electric fields generated by the transmission line interfered with the SET and prevented the application of greater RF powers.

We modified the pulse sequence to remove the electron before applying the RF excitation at the $\nu_{n0}$ transition (Fig. 4d). This enabled the demonstration of Rabi oscillations on the ionized (D$^+$) $^{31}$P nuclear spin (Figs. 4e,f). The Rabi oscillations now have near-unity visibility, because the electron spin state has no bearing on the nuclear resonance frequency while the donor is ionized.

**Nuclear spin qubit coherence times: Ramsey fringes and Hahn echo**

To assess the viability of utilizing the $^{31}$P nuclear spin as a quantum bit, it is critical to characterize the duration over which coherence is preserved. The time it takes a coherent superposition of nuclear spin states to evolve into an incoherent mixture, averaged over many experimental runs, is termed $T_2^*$. This important figure of merit can be found by carrying out a Ramsey fringe measurement, the NMR pulse sequence for which is shown in Fig. 5a. An initial $\pi/2$ pulse puts the nuclear spin in an equal superposition of $|\Uparrow\rangle$ and $|\Downarrow\rangle$, or equivalently, in the XY-plane on the Bloch sphere. We let it evolve for a time $\tau$ before executing another $\pi/2$ pulse and performing a measurement on the nuclear spin (see Fig. 5b for a Bloch sphere state evolution). We repeat the sequence 200 times and then step $\tau$, with the acquisition of each $\tau$ occurring over ~ 3 minutes. The spin is intentionally detuned from resonance so that during the period of free evolution, a phase is accumulated between the states $|\Uparrow\rangle$ and $|\Downarrow\rangle$. Consequently, interference fringes/oscillations are observed in the recovered nuclear spin flip probability as a function of $\tau$ (Fig. 5c). The decay of the fringes in Fig. 5c is the result of fluctuations in the local magnetic environment, which cause slight variations in the strength of detuning between runs. Fitting the data with a damped cosine function $P_n(\tau) = P_n(0)\cos(2\pi\Delta d\tau)\exp(-\tau/T_2^*)$, where $P_n(0)$ is the amplitude and $\Delta d$ the average detuning from resonance, reveals a $T_2^*(\text{D}^0) = 0.84(10)$ ms for the neutral donor and a $T_2^*(\text{D}^+) = 3.3(3)$ ms for the ionized donor. These figures are ~ $10^4$ times longer than those measured for the electron spin[18,19].

Many of the magnetic fluctuations that contribute to $T_2^*$ occur on timescales much greater than the typical nuclear spin manipulation time (~ 25 μs for a π pulse). Therefore, a significant portion of the dephasing can be reversed by performing a π rotation in the middle of the free evolution period of Fig. 5a. This modified sequence (Fig. 5d) is known as a Hahn echo (refer to Fig. 5e for a Bloch sphere representation). Observing the echo signal as the delay $\tau$ is varied yields the decay curves displayed in Fig. 5f. We fit the data with functions of the form $y = y(0)\exp((-2\tau/T_2)^b)$, where $y(0)$ is the amplitude, $b$ is a free exponent and $T_2$ is the coherence time. For the neutral donor spin, we find $T_2(\text{D}^0) = 3.5(1)$ ms and $b(\text{D}^0) = 2.2(2)$, and for the ionized donor spin we extract $T_2(\text{D}^+) = 60.0(9)$ ms and $b(\text{D}^+) = 1.77(7)$.

The observation that the ratio $T_2/T_2^*$ is not the same for the D$^0$ and D$^+$ charge states of the donor nuclear spin (~ 4 for the former and ~ 18 for the latter) suggests that the power spectral density of the decohering noise may have a different form (not just strength) for the two cases[47]. For the ionized donor spin, both the coherence time ($T_2(\text{D}^+) = 60$ ms) and the shape of the echo decay ($b(\text{D}^+) = 1.77$) are fully accounted for by the spectral diffusion caused by dynamics of the $^{29}$Si nuclear spin bath, as quantified by recent theory[48]. Accordingly, we expect that removal of $^{29}$Si through isotropic purification[28] should yield $T_2$ values of order minutes, as observed in bulk-doped samples[32].

For the neutral donor spin, one would expect the mechanism of $^{29}$Si spectral diffusion to be weaker than in the ionized case, as the hyperfine field gradient provided by the electron should, to some extent, suppress the $^{29}$Si dipole flip-flops[49]. The observation that the nuclear spin coherence is worse for the neutral donor suggests that an additional decoherence process occurs there. One possibility is that charge noise at the Si/SiO$_2$ interface[45] or electronic gate noise[27] causes a time-dependent Stark shift of the hyperfine coupling, which results in a random modulation of the instantaneous nuclear Larmor frequency. Future work will focus on the identification and mitigation of these additional processes.

**Qubit readout and control fidelities**

We now turn to an analysis of the fidelity of our solid-state qubit. The single-shot nuclear spin readout can be performed with a very high fidelity, owing to the quantum non-demolition (QND) nature of the measurement. In general, a QND measurement is obtained if the Hamiltonian $H_{\text{int}}$, describing the interaction between observable and measurement apparatus, commutes with the observable[50]. In our case, the "observable" is the $z$ projection of the nuclear spin state $I_z$, while the "measurement apparatus" is the electron spin. The QND condition, $[I_z, H_{\text{int}}] = 0$, would require a hyperfine coupling of the form $AS_z I_z$. The physical phenomena responsible for the observed nuclear spin quantum jumps originate from the measurement through the electron spin, and can be viewed as a deviation from QND ideality. The isotropic hyperfine coupling contains the terms $A_\parallel (S_x I_x + S_y I_y)$, which do not commute with $I_z$. In addition any anisotropic part of the hyperfine tensor $\underline{A}$ (e.g., $A_\perp S_z I_x$) also does not commute with $I_z$. Here, $A_\parallel$ ($A_\perp$) represents the diagonal (off-diagonal) components of the hyperfine tensor $\underline{A}$ (see Supplementary Information for details). For the nuclear $|\Uparrow\rangle$ state, this results in a lifetime $T_\Uparrow = 1495(360)$ s (obtained from extended data of the measurement in Fig. 2a). For the nuclear $|\Downarrow\rangle$ state, the cross-relaxation process - caused by phonons modulating the hyperfine coupling - also introduces a term that does not commute with $I_z$, yielding a lifetime $T_\Downarrow = 65(15)$ s (Fig. 2a).

These lifetimes must be contrasted with the nuclear spin measurement time $T_{\text{meas}}$, which has been optimized here to maximize the nuclear spin readout fidelity (refer to Methods Summary). Combining the optimal measurement time ($T_{\text{meas}} = 104$ ms) with the observed nuclear spin lifetimes yields the QND fidelities: $F_{\text{QND}}(|\Uparrow\rangle) = \exp(-T_{\text{meas}} / T_{\Uparrow}) = 0.99993(2)$; and $F_{\text{QND}}(|\Downarrow\rangle) = \exp(-T_{\text{meas}} / T_{\Downarrow}) = 0.9984(4)$. We have therefore obtained readout fidelities between 99.8% and 99.99%, the highest for any solid-state qubit, and comparable with the fidelities observed for qubits in vacuum-based ion-trap systems[11].

Next we explore the nuclear spin control fidelity. The rotation angle error $\theta$ can be estimated for both the neutral and ionized donor nuclear spins by simulating the Rabi oscillations assuming Gaussian fluctuations of their instantaneous resonance frequencies, a method previously adopted in ref. 19. We deduce the standard deviation of these fluctuations from the measured pure dephasing times $T_2^*$, and find a resulting best case 1-qubit gate ($\pi$ rotation) control fidelity $F_C = \left(1 - \theta/180°\right)$ of 99.988(2)% for the neutral donor and 99.9986(1)% for the ionized donor. Extrinsic sources of pulse error, for example due to power fluctuations of the RF source, are not captured in these simulations. We have measured the rotation angle error for the ionized donor nuclear spin using multiple-pulse dynamical-decoupling sequences (see Supplementary Information), and extracted a maximum uncertainty of 3° for an intended $\pi$ pulse, indicating a lower-bound on the control fidelity $F_C$ of 98%, in agreement with the Rabi oscillation simulations.

**Perspective**

The results presented here demonstrate the ability to combine electrical single-shot readout with coherent control of the nuclear spin of a single $^{31}$P donor. We have shown that the $^{31}$P qubit allows exquisite readout and control fidelities – on par with the best atomic systems in vacuum – while being hosted in a silicon chip, inserted by ion implantation, and operated electrically using nanostructures compatible with standard silicon-MOS fabrication. We anticipate that exploiting the $^{31}$P nuclear spin qubit will open new avenues to pursue large-scale quantum computer architectures, where the quantum coherence of well-isolated atomic systems is combined with the manufacturability of silicon nanoelectronic devices.

## METHODS SUMMARY

**Device fabrication.** For information relating to the device fabrication we refer the reader to ref. 19, where it has been described in some detail.

**Experimental setup.** For our voltage pulses, we employed a compensation technique using a Tektronix AWG520, to ensure that the pulsing only shifted the donor electrochemical potentials but kept the SET island potential constant. The voltage $V_p$ (see Supplementary Information Fig. S1b) was applied directly to the top gate, while it was inverted and amplified by a factor $K$ before reaching the plunger gate. The gain $K$ was carefully tuned to ensure that the SET operating point moved along the top of the SET current peaks, as shown by the blue arrow in Fig. 1d of ref. 36. The SET current was measured by a Femto DLPCA-200 transimpedance amplifier at room temperature, followed by a voltage post-amplifier, a 6th order low-pass Bessel filter, and a fast digitising oscilloscope.

The ESR excitations were produced by an Agilent E8257D microwave analog signal generator and the NMR excitations by an Agilent MXG N5182A RF vector signal generator. The two signals were combined at room-temperature with a power divider/combiner, before being guided to the sample by a semi-rigid coaxial cable (2.2 m in length with a loss of ~ 50 dB at 50 GHz). Gating of the ESR/NMR pulses was provided by the Tektronix AWG520, which was synchronized with the TG and PL pulses. For the nuclear spin readout, adiabatic inversion of the electron spin was achieved by applying frequency chirps at a rate of 50 kHz and with a peak-to-peak deviation of 20 MHz.

**Optimizing the nuclear spin readout time.** The nuclear spin readout fidelity depends on the measurement time $T_{\text{meas}}$, which is given by the number of single-shot electron spin readout events acquired per measurement. The larger the number of electron spin single-shots taken, the lower the SNR-limited readout error will be. However, a trade-off exists in that the longer measurement duration results in a greater chance of a nuclear spin quantum jump occurring during readout. In Figs. 2a-e, 250 shots resulted in a 260 ms measurement duration and a SNR-limited readout error of $2\times10^{-7}$ (see above section Electrical single-shot nuclear spin readout). By decreasing the number of shots to 100, we were able to perform the measurement in less than half of the time ($T_{\text{meas}} = 104$ ms) with an increased, but still relatively low SNR-limited error of $2\times10^{-5}$. We find that 100 shots provides a near optimal tradeoff between nuclear spin quantum jump occurrence and the SNR-limited readout error.

**Acknowledgements** We thank R.P. Starrett, D. Barber, C.Y. Yang and R. Szymanski for technical assistance and A. Laucht, R. Kalra and J. Muhonen for fruitful discussions. This research was funded by the Australian Research Council Centre of Excellence for Quantum Computation and Communication Technology (project number CE11E0096) and the US Army Research Office (W911NF-08-1-0527). We acknowledge support from the Australian National Fabrication Facility.

**Author contributions** K.Y.T. and W.H.L. fabricated the device; D.N.J. designed the P implantation experiments; J.J.P., K.Y.T., J.J.L.M., J.P.D. and F.A.Z. performed the measurements; J.J.P., A.M., A.S.D. and J.J.L.M. designed the experiments and discussed the results; J.J.P. and A.M. analysed the data; J.J.P. and A.M. wrote the manuscript with input from all coauthors.

Correspondence should be addressed to A.M. (a.morello@unsw.edu.au).

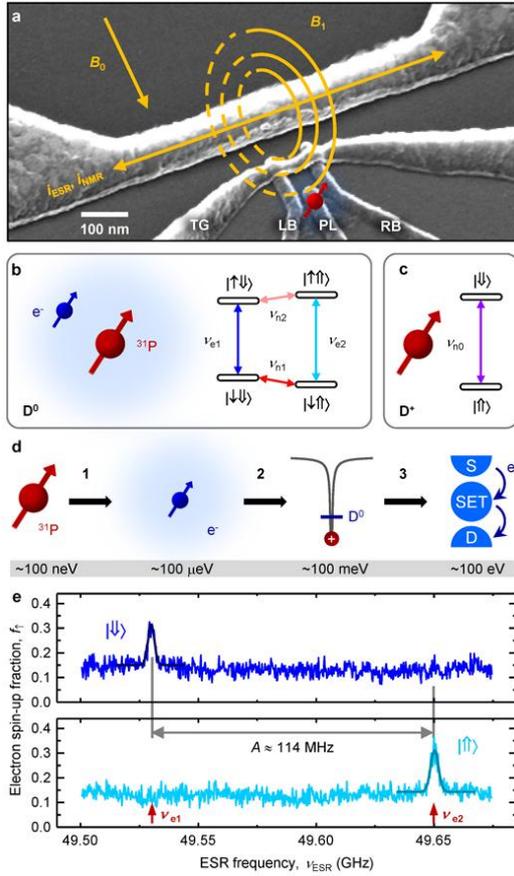

**Figure 1 | Qubit nanostructure and spin transitions. a,** Scanning electron micrograph of the active area of the qubit device, showing an implanted donor (donor as red arrow), the SET and the short-circuit termination of the microwave line. The device is mounted in a dilution refrigerator with electron temperature ~ 300 mK, and subject to static magnetic fields between 1.0 and 1.8 T. **b,** Energy level diagram of the neutral $^{31}$P donor system, with corresponding ESR (blue) and NMR (red) transitions. ↓↑: electron spin states; ⇓⇑: nuclear spin states. **c,** Energy level diagram of the ionized $^{31}$P donor, with the single NMR transition shown in purple. **d,** $^{31}$P nuclear spin readout through our three stage amplification process. (1) The nuclear spin state (~ 100 neV energy splitting) is mapped onto the electron spin state (~ 100 μeV energy splitting) by performing ESR. (2) The electron spin state is projected onto a charge state of the donor (~ 50 meV) by using spin-to-charge conversion. (3) The SET provides a current that is dependent on the donor charge state (total signal energy of ~ 100 eV), thus revealing the nuclear spin state. **e,** ESR spectra obtained at $B_0 = 1.77$ T by scanning the microwave frequency and monitoring the electron spin-up fraction $f_↑$. The top trace corresponds to an active $\nu_{e1}$ ESR transition (nuclear spin state $|⇓⟩$) and the bottom trace to an active $\nu_{e2}$ ESR transition (nuclear spin state $|⇑⟩$).

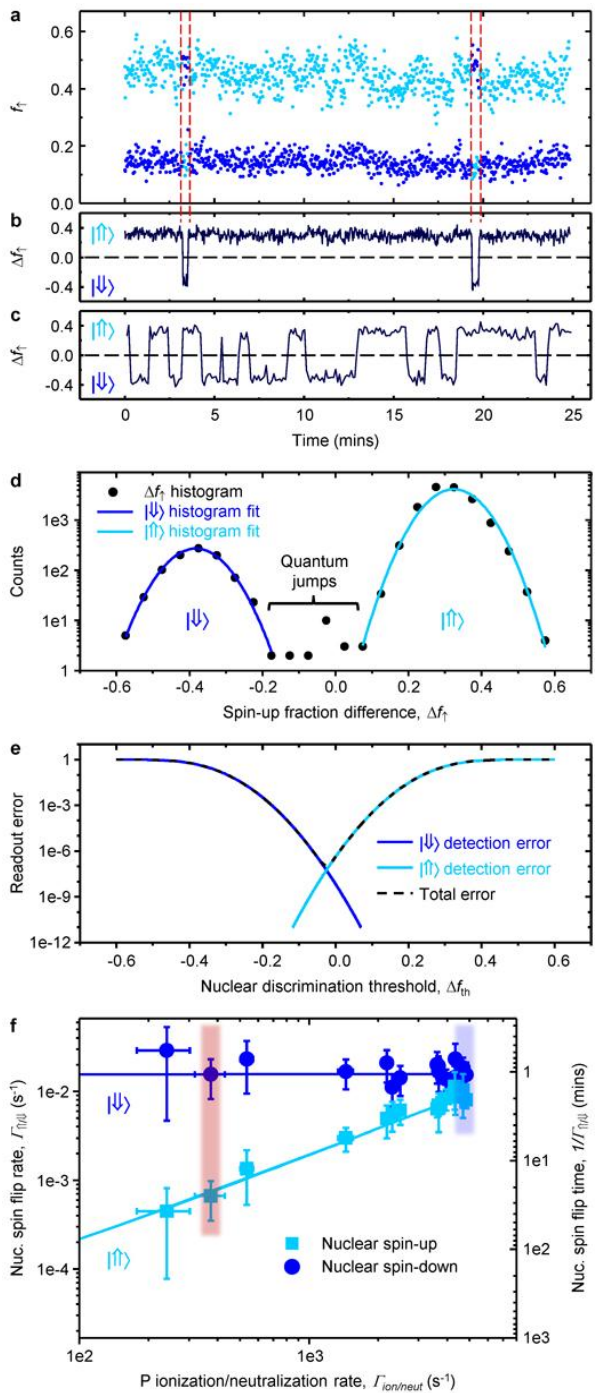

**Figure 2 | Nuclear spin quantum jumps, lifetimes and readout error. a,** Repetitive single-shot measurements of the nuclear spin state, obtained by toggling $\nu_{ESR}$ between $\nu_{e1}$ = 49.5305 GHz (dark blue) and $\nu_{e2}$ = 49.6445 GHz (light blue), and recording the electron spin-up fraction $f_\uparrow$. Each data point represents the average $f_\uparrow$ over 250 measurements. **b,** Electron spin-up fraction difference, $\Delta f_\uparrow = f_\uparrow(\nu_{e2}) - f_\uparrow(\nu_{e1})$, for the data in panel **a**. $\Delta f_\uparrow > 0$ indicates nuclear spin $|\Uparrow\rangle$, and vice versa. **c,** $\Delta f_\uparrow$ in an experiment with an additional resonant tunneling phase, to enhance the electron ionization/neutralization rate (see text). **d,** Histograms of $\Delta f_\uparrow$ for the data in panel **b**, showing two well-separated Gaussian peaks, each corresponding to a nuclear spin state as indicated. The counts obtained for $-0.015 < \Delta f_\uparrow < 0.05$ are attributed to nuclear spin quantum jumps occurring during the measurement. **e,** Readout errors as a function of the detection threshold for $\Delta f_\uparrow$. **f,** Nuclear spin flip rates $\Gamma_{\Uparrow/\Downarrow}$ as a function of the donor ionization/neutralization rate $\Gamma_{ion/neut}$. The light blue line is a fit to $\Gamma_\Uparrow = \Gamma_0 + p\Gamma_{ion/neut}$, with $p = 1.91(8)\times 10^{-6}$. The dark blue line is a constant $\Gamma_\Downarrow = 1.54(17)\times 10^{-2}$ s$^{-1}$. The red and blue shading indicate the values obtained from the data sets in panels **b** and **c**, respectively. Calculation of the error bars is described in the Supplementary Information.

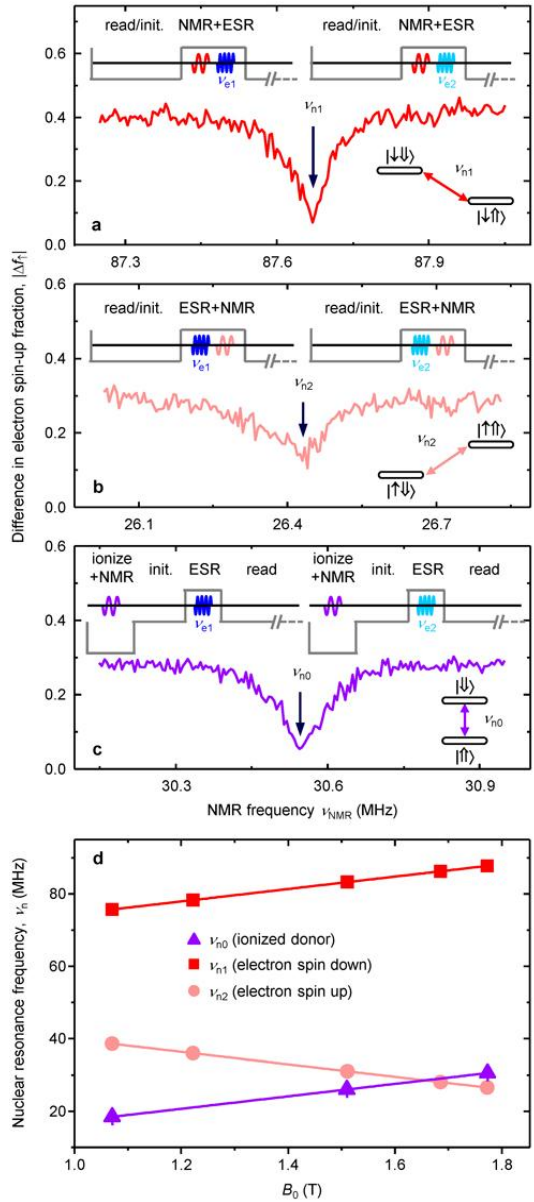

**Figure 3 | Nuclear magnetic resonance of a single $^{31}$P nucleus. a-c,** Observation of nuclear resonances at $B_0 = 1.77$ T, while the electron spin is $|\downarrow\rangle$ (**a**), $|\uparrow\rangle$ (**b**), or absent, i.e. ionized donor (**c**). The resonance condition is obtained when $|\Delta f_\uparrow|$ drops from the unperturbed value $\approx 0.4$ to near zero, due to the randomization of the nuclear spin state. In each panel, the top inset shows the plunger gate voltage waveform (grey line) plus NMR/ESR pulse sequence, whilst the bottom-right inset shows the energy levels involved in the NMR transition. **d,** Dependence of the NMR resonances on the magnetic field $B_0$. Solid lines are the values predicted using the $^{31}$P nuclear gyromagnetic ratio $\gamma_n = 17.23$ MHz/T.

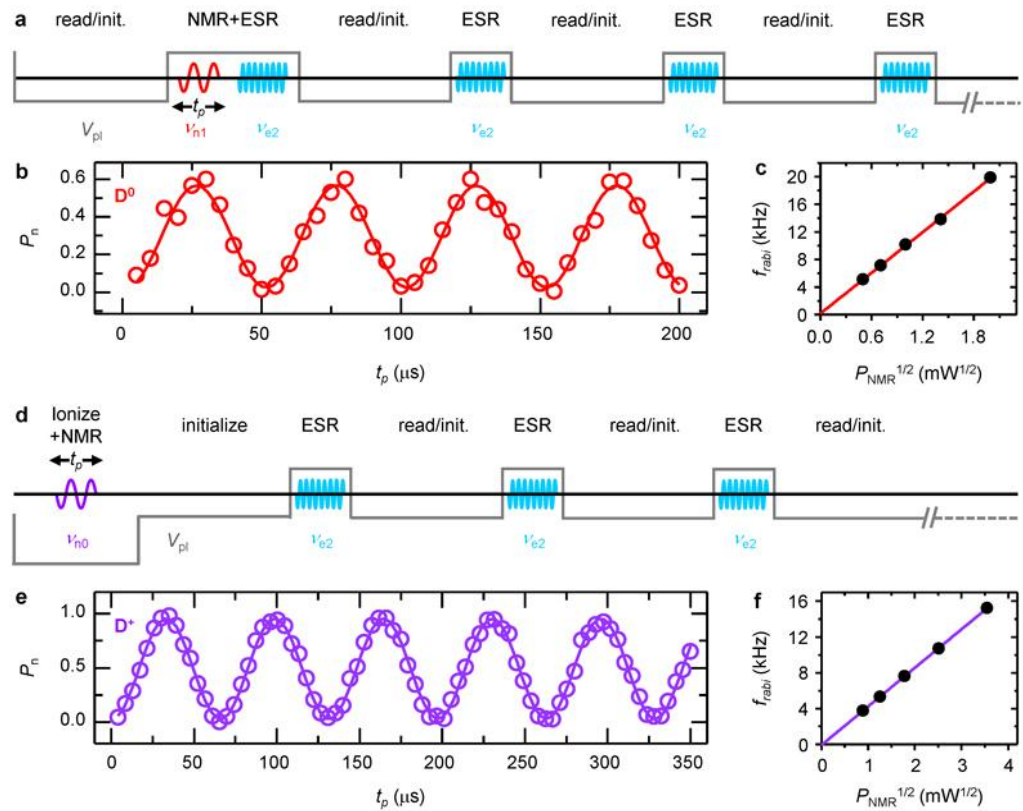

**Figure 4 | Rabi oscillations of a single $^{31}$P nuclear spin. a,** Pulse sequence for the coherent rotation of a $^{31}$P nuclear spin with the donor in the neutral D$^0$ state. Depicted is the plunger gate voltage waveform (grey line) and combined NMR/ESR pulses. After a coherent NMR pulse at $\nu_{n1}$ of duration $t_p$, the state of the nuclear spin is read by probing the $\nu_{e2}$ ESR transition with 300 single-shot adiabatic inversion and electron spin readout measurements, lasting approximately 300 ms. We note that we could have equally chosen to perform the readout at the $\nu_{e1}$ transition. The resulting electron spin-up fraction $f_\uparrow(\nu_{e2})$ is compared to a threshold, extracted from the quantum jumps experiment (Fig. 2a), and a nuclear spin orientation is ascribed to the measurement. **b,** Rabi oscillation of the neutral $^{31}$P donor nuclear spin with $P_{NMR}$ = 6 dBm, $B_0$ = 1.07 T and $\nu_{n1}$ = 75.7261 MHz. The pulse sequence of **a** is repeated 40 times for each Rabi pulse length $t_p$, with 5 sweeps of $t_p$ performed to give a total of 200 measurements at each $t_p$. The number of nuclear spin flips is recorded to give the flip probability $P_n$. The solid line is a fit of the form $P_n = K\sin^2(\pi f_{rabi} t_p)$, where $K$ and $f_{rabi}$ are free fitting parameters. **c,** Rabi frequency $f_{rabi}$, extracted from fits of data similar to that in panel **b**, against the RF excitation amplitude $P_{NMR}^{1/2}$. **d,** Modified pulse sequence to perform Rabi oscillations on the $^{31}$P nuclear spin with the donor in the ionized D$^+$ state. The electron is first removed before a coherent NMR burst is applied. The electron is then replaced so that a single-shot measurement can be performed on the nuclear spin. **e,** Sample Rabi oscillation of the ionized donor nuclear spin using $P_{NMR}$ = 21 dBm, $B_0$ = 1.77 T and $\nu_{n0}$ = 30.5485 MHz, with each data point again comprising 200 nuclear spin state measurements. **f,** Plot showing the linear scaling of the ionized nuclear spin Rabi frequency with the excitation amplitude.

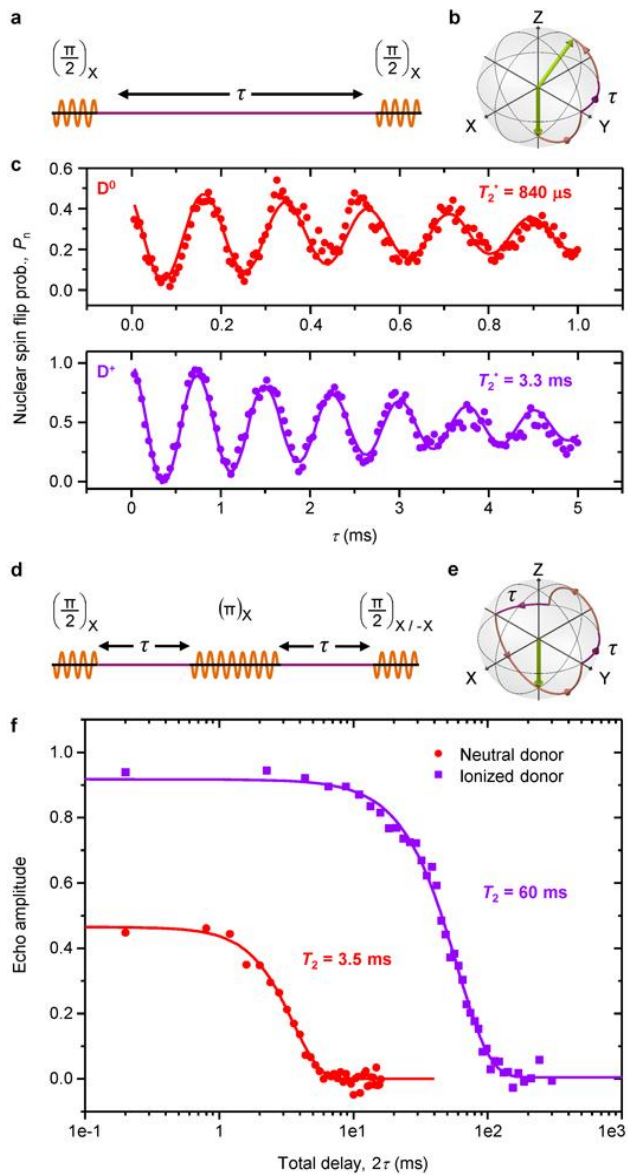

**Figure 5 | Ramsey fringes and spin echo decay. a,** NMR pulse sequence for the Ramsey fringe experiment. This sequence replaces the single pulse of duration $t_p$ in Fig. 4a (Fig. 4d) for the neutral (ionized) donor, whereas the nuclear spin is read out in the same way. The phase of both $\pi/2$ pulses is such that rotation is performed about the X axis on the Bloch sphere, as noted in the rotation angle subscript above each pulse. **b,** Bloch sphere representation of the evolution in the rotating frame for the Ramsey fringe measurement. The green arrow represents the nuclear spin. The purple path represents dephasing in between pulses, whilst the orange path represents a rotation about X. **c,** Ramsey interference fringes for the nuclear spin with the donor in the $D^0$ (top) and $D^+$ (bottom) charge states, taken at $B_0 = 1.77$ T. Here a $\pi/2$ pulse was 12.5 μs for the $D^0$ experiment and 23.5 μs for the $D^+$. We sweep the inter-pulse delay, and repeat the sequence 20 times at each $\tau$. A total of 10 sweeps are performed (200 measurements) and the nuclear spin flip probability $P_n$ is found. Fits to the data are discussed in the main text. **d,** Pulse sequence for the Hahn echo experiment. The full measurement protocol is provided by inserting this sequence in place of the single NMR pulse in Fig. 4a (Fig. 4d) for the neutral (ionized) donor. Here we also implement phase cycling, where the final $\pi/2$ rotation is first performed about the X axis and then the measurement is repeated with the final $\pi/2$ about the -X axis. Subtracting the two signals ensures a baseline of zero. **e,** Bloch sphere representation for the Hahn echo measurement. Here the final $\pi/2$ pulse is about X (-X is not shown). **f,** Decay of the echo amplitude as the delay $\tau$ is increased for the case of a neutral (circles) and ionized (squares) donor. We perform 40 repetitions of the sequence for each $\tau$ and 25 sweeps, totalling 1000 measurements at each point. The phase-cycled echo amplitude is given by $[P_n(\nu_n, -X) - P_n(\nu_n, X)]/40$, where $P_n(\nu_n, -X / X)$ represents the nuclear spin flip probability measured at the NMR resonance $\nu_n$ with a final $\pi/2$ pulse about the -X or X axis. All other experimental conditions are as in the Ramsey fringe experiment. Fits through the data are discussed in the text.